\documentstyle[aaspp4]{article}

\def\ltsima{$\; \buildrel < \over \sim \;$}
\def\simlt{\lower.5ex\hbox{\ltsima}}            
\def\gtsima{$\; \buildrel > \over \sim \;$}
\def\simgt{\lower.5ex\hbox{\gtsima}}            

\newcommand{\asca}{{\it ASCA}}
\newcommand{\rosat}{{\it ROSAT}}
\newcommand{\sax}{{\it BeppoSAX}}
\newcommand{\einstein}{{\it Einstein}}
\newcommand{\ginga}{{\it Ginga}}
\newcommand{\heao}{{\it HEAO1}}

\newcommand{\logn}{Log $N$ - Log $S$ relation}
\newcommand{\Logn}{Log $N$ - Log $S$ Relation}
\newcommand{\etal}{{\it et al.}}

\newcommand{\ergs}{erg s$^{-1}$ cm$^{-2}$}

\newcommand{\de}{deg$^2$}

\lefthead{Ueda et al.}
\righthead{ASCA Medium Sensitivity Survey}

\begin{document}

\title{
The \asca\ Medium Sensitivity Survey (the GIS Catalog Project): 
Source Counts and Evidence for Emerging Population of Hard Sources}
\author{
Yoshihiro Ueda\altaffilmark{1},
Tadayuki Takahashi\altaffilmark{1},
Yoshitaka Ishisaki\altaffilmark{2},
Takaya Ohashi\altaffilmark{2}, \\
and Kazuo Makishima\altaffilmark{3}
}

\altaffiltext{1}{Institute of Space and Astronautical Science, Yoshinodai 3-1-1, Sagamihara, Kanagawa 229-8510, Japan}
\altaffiltext{2}{Department of Physics, Tokyo Metropolitan University, Hachioji, Tokyo 192-0397, Japan}
\altaffiltext{3}{Department of Physics, University of Tokyo, Bunkyo-ku, Tokyo 113, Japan}

\begin{abstract}

We present first results from the \asca\ Medium Sensitivity Survey
(AMSS; or the GIS catalog project). From the serendipitous fields
amounting to 106 deg$^2$, we determined the \logn s in the 0.7--7 keV
and 2--10 keV bands with the best statistical accuracy obtained so
far, over the flux range from $1\times10^{-11}$ to $5\times10^{-14}$
and $7\times10^{-14}$ \ergs , respectively. When the sources detected
in the 0.7--7 keV band are divided into two subsamples with higher and
lower spectral hardness, the former exhibits a significantly steeper
slope than the latter at fluxes below $\sim 10^{-12}$ \ergs\ (0.7--7
keV). The average spectrum of sources becomes continuously harder
toward fainter fluxes, from a photon index of 2.1 in the 0.7--10 keV
range at the flux of $\sim 10^{-11}$ to 1.6 at $\sim 10^{-13}$ \ergs\
(0.7--7 keV). This is consistent with the comparison of source counts
between the 2--10 keV and the 0.7--2 keV band, and solves the puzzle
of their discrepancy reported previously. Our results demonstrate
rapid emergence of hard X-ray sources with a decreasing flux from
$\sim10^{-12}$ to $\sim10^{-13}$ \ergs\ (2--10 keV).

\end{abstract}



\section{Introduction}

X-ray surveys are the most direct approach to reveal the nature of
sources that contribute to the Cosmic X-ray Background (CXB or XRB;
see Fabian and Barcons 1989 for review). The \rosat\ satellite
resolved 70--80\% of the CXB in the 0.5--2 keV band into discrete
sources, whose majority are Active Galactic Nucleis (AGNs) (e.g.,
Hasinger \etal\ 1998; Schmidt \etal\ 1998). Because of the technical
difficulties, imaging sky surveys in the hard X-ray band (above 2
keV), where the bulk of the CXB emission arises, were not available
until the launch of \asca\ (Tanaka, Inoue, and Holt, 1993). 
Deep/medium surveys performed with \asca\ (Ogasaka \etal\ 1998; Ueda
\etal\ 1998; 1999) and \sax\ (Fiore \etal\ 1999) have resolved a
significant fraction (25--35\%) of the CXB above 2 keV. The results
indicate emergence of faint, hard X-ray sources that could be
responsible in producing the CXB spectrum, which is harder than that
of nearby type-I AGNs.

These surveys, though novel, are limited in sky coverage. As a result,
the sample size of detected sources is not sufficient to obtain a
self-consistent picture about the evolution of the sources over the
wide fluxes, from $\sim10^{-11}$ \ergs\ (2--10 keV) which is the
sensitivity limit of \heao\ A2 (Piccinotti
\etal\ 1982), down to $\sim 10^{-13}$ \ergs\ (2--10 keV), 
that of \asca\ (e.g., Ueda \etal\ 1999). In particular, the broad band
properties of sources at these fluxes are somewhat puzzling according
to previous studies. The source counts in the soft band (0.3--3.5 keV)
obtained by \einstein\ Extended Medium Sensitivity Survey (EMSS; Gioia
\etal\ 1990) is about 2 times smaller than that in the hard band
(2--10 keV) obtained by the \ginga\ fluctuation analysis (Hayashida,
Inoue, \& Kii 1990; Butcher \etal\ 1997) when we assume a power-law
photon index of 1.7. This may imply presence of many hard sources at
the flux level of $\sim 10^{-12}$ \ergs\ (2--10 keV). Such evidence is
not seen, however, in the spectrum of the fluctuation observed by
\ginga, which shows a photon index of $1.8\pm0.1$ in the 2--10 keV
range (Butcher \etal\ 1997).

To complement these shortcomings of deep surveys, we have been working
on the project called the ``\asca\ Medium Sensitivity Survey (AMSS)'',
or the GIS catalog project (Ishisaki \etal\ 1995; Ueda \etal\ 1997;
Takahashi \etal\ 1998). In the project, we utilize the GIS data from
the fields that have become publicly available to search for
serendipitous sources. The large field of view and the low-background
characteristics make the GIS instrument ideal for this purpose (Ohashi
\etal\ 1996; Makishima \etal\ 1996).
In this letter, we present the first results obtained from the latest
version of the GIS catalog, in which the data taken from May 1993
through December 1996 are analyzed. The source list and detailed
description of the catalog will be presented in a separate paper. From
the serendipitous fields amounting to 106 \de, we derived the \logn s
in the 0.7--7 keV and 2--10 keV bands with the best statistical
accuracy obtained so far, over the wide flux range of more than 2
decades. The AMSS sample, currently the largest sample in the 0.7--10
keV range, leads us to the best understanding of the statistical
properties of sources that produce about 30\% of the CXB.

\section{Results}

\subsection{Field Selection and Data Used in the Analysis}

The first \asca\ GIS source catalog (Ueda \etal\ 1999, in preparation)
contains 1345 sources, including ``target sources'' at which the
observation is aimed, detected from 369 fields observed from 1993 May
to 1996 December. The analysis method for source surveys is basically
the same as applied for the \asca\ LSS (Ueda \etal\ 1998; 1999) except
that the SIS data are not used in the AMSS. In the procedure we took
into account the complicated responses of the XRT and the detectors. 
The analysis procedure consists of two steps: {\bf I. source
detection} and {\bf II. flux calculation}. In step~I, raw images are
cross-correlated (smoothed) with the position dependent PSF of the XRT
and the GIS and source candidates are searched in the smoothed image. 
In step~II, we perform a 2-dimensional maximum-likelihood fitting to
the raw image with a model that consists of the background and source
peaks found in step~I, including that of a target source. Multiple
observations of the same or overlapped fields are combined in the
analysis. The following selection criteria are applied for the
catalog: (1) the Galactic latitude $|b|$ is higher than $10^\circ$,
(2) the time-averaged count rate is less than $0.8$ c s$^{-1}$ per
sensor, (3) the exposure is longer than 5000 sec, and (4) the
2-dimensional fit in step~II is successful (reduced $\chi^2 < 1.7$). 
We also excluded any fields for which we could not model the surface
profile adequately, such as the fields of bright cluster of galaxies.

In this paper, to limit our interest on the study of extra-galactic
sources, we further selected the fields of $|b| > 20^\circ$, excluding
observations of nearby galaxies of large angular size, star forming
regions, and supernova remnants. Finally, we discarded the target
sources from the source list to construct a complete sample consisting
of only serendipitous sources. The sample contains 714 sources (above
5 $\sigma$ detection), of which 696, 323, and 438 sources are detected
in the 0.7--7 keV (total), 2--10 keV (hard), and 0.7--2 keV (soft)
band, respectively. The number of sources detected both in the total
and hard bands, total and soft bands, and hard and soft bands are 320,
423, and 266, respectively, and 266 sources are detected in all the
survey bands. The total sky area covered amounts to 106 deg$^{-2}$ and
the sensitivity limits are $5\times10^{-14}$, $7\times10^{-14}$, and
$2.6\times10^{-14}$ \ergs\ for the 0.7--7 keV, 2--10 keV, and 0.7--2
keV survey band, respectively. For each field, we corrected the count
rates for the Galactic absorption, whose column density is estimated
from H~I observations (Dickey \& Lockman 1990). The LSS sources (Ueda
\etal\ 1999) are not included in the sample, whereas most of the GIS
data of the deep surveys are included there.

\subsection{The Source Spectra}

Figure~1 shows the correlation between the 0.7--7 keV flux and the
hardness ratio in the 0.7--10 keV range, $HR1$, defined as
$(H-S)/(H+S)$, where $H$ and $S$ represent the count rate within
radius of $5'.9$ around the source (corrected for vignetting, see Ueda
\etal\ 1999) in the 2--10 keV and 0.7--2 keV bands, respectively. Due
to the limited photon statistics, however, these spectral information
of individual sources is subject to a large statistical error,
typically by 0.03, 0.1, and 0.2 (1$\sigma$) at fluxes of $10^{-11}$,
$10^{-12}$, and $10^{-13}$ \ergs , respectively. Hence, to study the
average properties of the source spectra, we calculated the average
value of $HR1$ (weighting with their errors) in several flux ranges
from sources detected in the total band. The results are plotted in
Figure~1 with crosses.  Since the sensitivity is given in count rate
rather than in flux, we sort sources by count rate (not by flux), to
avoid to introduce any selection effects that very hard (or soft)
sources are difficult to detect at the flux close to the sensitivity
limit of each observation. The dashed curves in Figure~1 show conditions
that give the same count rate, and sources located in the regions
between the two curves are used for calculation of average hardness
ratio. It is clear from Figure~1 that the average spectrum becomes
harder with a decreasing flux. The corresponding photon index
(assuming a power law over the 0.7--10 keV band with no absorption)
changes from 2.1 at the flux of $\sim 10^{-11}$ \ergs\ to 1.6 at $\sim
10^{-13}$ \ergs\ (0.7--7 keV).

\subsection{\Logn}

Using the sample, we derived the \logn\ independently in the three
survey bands, 0.7--7 keV, 2--10 keV, and 0.7--2 keV, in the following
manner. For each field, we calculated the observed area as a function
of count rate, $\Omega(S)$, following the same procedure described in
Ueda \etal\ (1999). In this process, we estimate the significance of
detection expected for the given flux at every positions, utilizing
the fitting model used in step~II of the source finding procedure and
taking into account the presence of the target source. In calculation
of $\Omega(S)$ we excluded the regions where the spill of the PSF from
the target source is higher than 50\% of the background, and did not
count any sources detected within the excluded region. Then, after
summing up from all the selected fields $\Omega(S)$, and $N(S)$, the
number of sources in flux bin of $dS$, we calculated the ``observed''
\logn\ in the differential form by dividing $N(S)$ by $\Omega(S)$.

To evaluate possible systematic errors in the observed \logn s, we
next performed a large number of Monte Carlo simulations with roughly
the same exposure distribution as the real data: we created 1400
simulated image data, consisting of 400 pointings with an exposure of
20 ksec, 300 with 30 ksec, 500 with 40 ksec, 50 with 70 ksec, and 50
with 90 ksec. The \logn\ input to the simulations is taken to be the
same as the one we derived from the AMSS. We then applied the same
analysis procedure for these simulated data as for the real data, and
compared the derived parameters with the input parameters. We found an
excess of the output source counts relative to the input ones, which
is about 10\% (20\%) at $S=2\times10^{-13}$ ($8\times10^{-14}$) \ergs\
and 20\% (40\%) at $S=7\times10^{-14}$ ($3\times10^{-14}$) \ergs\ in
the total (soft) band, while such deviation is negligible in the hard
band. The deviation and its energy dependence are also seen in the
case of the LSS (Ueda \etal\ 1999) and can be explained by the effect
of the source confusion, because the positional resolution of the GIS
becomes worse toward lower energies. Thus, we corrected the observed
differential \logn\ for this effect with a multiplicative factor $f$,
derived by the simulation, which has a form of $ f = 1 / [\Delta
\times ({\rm log}S-{\rm log}S_0) / ({\rm log}S_1-{\rm log}S_0) + 1]$
for $S<S_0$, and f = 1 for $S>S_0$, where $\Delta$ = 0.2 (0.4), $S_0 =
5.9\times10^{-13} (2.2\times10^{-13})$, and $S_1 = 7.4\times10^{-14}
(2.7\times10^{-14})$, for the total (soft) band survey, respectively. 
These simulated data are also used to evaluate the systematic error
due to the source confusion in determination of $HR1$: the amount of
the bias is estimated to be $-0.03$ at $10^{-13}$ \ergs\ (0.7--7 keV)
and is negligibly small in brighter flux level. This bias has been
taken into account in deriving the average hardness ratio in Figure~1.

Since we impose the maximum count rate in the field-selection
criteria, very bright sources are intentionally excluded in our
sample: this means there is an upper flux limit above which the sample
becomes incomplete. The maximum count rate, 0.8 c/s/sensor including
the background and the target sources, indicates the upper flux limit
is higher than $(1\sim1.5) \times 10^{-11}$ \ergs\ (2--10 keV) for
most of the fields, although it depends both on the spectrum and the
brightness of the target source. In integrating the differential \logn
s, we set the upper flux limits at $10^{-11}$ \ergs\ for the 0.7--7
keV and 2--10 keV surveys, and $5\times 10^{-12}$ \ergs\ for the 0.7--2 keV
survey, and used previous results by
\heao\ A2 (Piccinotti \etal\ 1982) as integral constants at the bright
flux ends assuming a photon index of 1.7. The uncertainty in the
completeness close to these limits does not affect our discussion
below, which are made mainly based on the fluxes fainter than
$10^{-12}$ \ergs\ (0.7--7 keV and 2--10 keV).

The large sample size enables us to derive \logn s separately for
sources with different spectra. In practice, considering the large
statistical error in the hardness ratio for an individual source, we
divide the sample into two, the ``soft source sample'', consisting of
sources with $HR1$ smaller than --0.028, which corresponds to a photon
index larger than 1.7, and the ``hard source sample'', with a photon
index smaller than 1.7. In the conversion from count rate into flux,
we assumed a photon index of 1.6 (1.6) for the hard source sample and
1.9 (1.6) for the soft source sample in the 0.7--7 keV (2--10 keV)
survey. These photon indices come from the average spectrum of each
sample after correcting for biases of purely statistical origin, based
on the simulation.

Figure~2 shows the integral \logn s in the 0.7--7 keV survey band for
the soft source sample (red curve), the hard source sample
(blue curve), and the sum (black curve). The figure
clearly demonstrates that sources with hard energy spectra in the
0.7--10 keV range are rapidly increasing with decreasing fluxes,
compared with softer sources. The ratio of the hard source sample
(with a photon index of less than 1.7) to the soft source sample
changes from about 20\% at the flux of $10^{-12}$ \ergs\ (0.7--7 keV)
to about 50\% at $10^{-13}$ \ergs . We confirmed by the simulations
that any statistical biases does not produce such dramatic change as a
function of flux. For the 2--10 keV survey, we show the sum of the
soft and hard source samples in Figure~3 (black curve).

\section{Discussion}

First, for confirmation, we compared the \logn\ in the 0.7--2 keV band
with the results by \rosat\ in almost the same band (0.5--2 keV)
(Hasinger 1998 and references therein). We found a good agreement
within the statistical error at $S > 1\times 10^{-13}$ \ergs\ (0.7--2
keV) between the two, whereas we see a slight excess of the
\asca\ source counts by about 10--20\% at $S =
3\times 10^{-14} \sim 1\times 10^{-13}$ \ergs\ (0.7--2 keV), which was
also reported from the LSS (Ueda \etal\ 1999). Using the AMSS data, we
found that these excess can be explained by contribution of
hard sources that have an (apparent) photon index less than 1.7 in the
0.7--10 keV range, whose number density increases more rapidly than
that of softer sources toward fainter fluxes in the 0.7--2 keV band. 
Thus, as is discussed in Ueda \etal\ (1999), it is plausible that the
difference between the \asca\ and \rosat\ results can be accounted for
by the presence of hard (or absorbed) sources, which are more easily
detectable by \asca\ than by \rosat, due to their difference of energy
dependence of the effective area even within the similar energy band.

The present results of the \logn s in the three survey bands are all
consistent with the LSS results (Ueda \etal\ 1999) within the
statistical errors. The 2--10 keV source counts is also consistent with
that by Cagnoni, Della Ceca, \& Maccacaro (1998), who utilized the
GIS2 data from about one fifth of the data set analyzed here, and with
the fluctuation analysis of the \asca\ SIS data by Gendreau, Barcons
\& Fabian (1998). In Figure~3, we show the constraints from the
fluctuation analysis by \ginga\ (Butcher \etal\ 1997). As noticed,
below the flux of $10^{-12}$ \ergs\ (2--10 keV), our direct source
counts gives somewhat lower values than \ginga, although it is barely
consistent within the 90\% statistical errors. 

The increasing fraction of hard sources toward fainter fluxes seen in
Figure~2 accounts for the evolution of the average source spectra
shown in Figure~1. The spectral evolution is also consistent with the
comparison of the source counts between the hard and the soft band. In
Figure~3, we compare the source counts in the hard band with that in
the soft band, both obtained from the AMSS. In the figure, the 0.7--2
keV fluxes are converted into the 2--10 keV fluxes assuming the two
photon indices (1.6 and 1.9). As is clearly seen, the source counts in
the hard band survey increases more rapidly than that in the soft band
survey toward the sensitivity limit. As a result, the hard band source
counts at $S\sim 10^{-13}$ \ergs\ (2--10 keV) matches the soft band
one when we assume a photon index of 1.6, whereas at brighter level of
$S = 4\times 10^{-13} \sim 10^{-12}$ \ergs\ (2--10 keV), we have to
use a photon index of about 1.9 to make them match. These indices,
which should represent average source spectra at each flux level, well
coincide with those directly derived from Figure~1. It is also
interesting that the slope of the \logn\ in the hard- and soft-source
sample in the total band survey (Figure~2) is similar to that of the
\logn\ in the hard- and soft-band survey (Figure~3), respectively, at
the flux range of one decade above the sensitivity limit. This is
expected if we roughly consider that the sources in the hard (soft)
source sample correspond to those detected in the hard (soft) survey.

As discussed above, all the present results from the AMSS are
perfectly consistent with one another. Consequently, the derived \logn
s covering the 0.7--10 keV range, determined by direct source counts,
has now solved the puzzle of discrepancy of the source counts between
the soft (EMSS) and the hard band (\ginga\ and \heao ). They are now
reconciled by the two facts: (1) the source counts in the hard band
obtained by the AMSS is smaller than the best-fit value of the \ginga\
results at $S = 4\times 10^{-13} \sim 1\times10^{-12}$ \ergs\ (2--10
keV) and (2) that in the soft band, which is consistent with the
recent \rosat\ surveys such as RIXOS (Mason \etal\ 1999), gives larger
source counts by $\sim$30\% than the EMSS results that include
Galactic objects. As seen from Figure~1, the average spectrum of
sources at $S = 3\times10^{-13} \sim 3\times10^{-12} $ \ergs\ (0.7--7
keV) has a photon index of 1.8--2.1 in the 0.7--10 keV range,
indicating that the contribution of hard sources (such as heavily
absorbed AGNs) is not significant yet at this flux level to reproduce
the CXB spectrum. This fact can be connected with the ``soft''
spectrum of the fluctuation observed with \ginga, which shows a photon
index of 1.8$\pm0.1$ in the 2--10 keV range (Butcher
\etal\ 1997).

The emerging population of hard sources seen in Figure~2 can be
understood by increasing contribution of absorbed AGNs, which should
become more significant toward fainter flux level due to the
K-correction effect as is discussed in the AGN synthesis model (Awaki
\etal\ 1991; Comastri \etal\ 1995). In fact, optical identification of
the \asca\ LSS shows that most of the hard (or absorbed) sources are
narrow-line or weak broad-line AGNs, and that their contribution is
comparable to that of unabsorbed AGNs at the 2--10 keV flux of
$2\times10^{-13}$ \ergs\ (Akiyama \etal\ 1999). Since the number of
sources in the LSS sample is limited, the optical identification of
the AMSS sample is crucial to reveal the evolution of each population,
particularly that of absorbed AGNs, which are difficult to detect in
the soft band. The integrated spectrum of sources with fluxes above
$\sim 10^{-13}$ \ergs\ (2--10 keV) is not hard enough to completely
account for the CXB spectrum. Our results thus predict that further
hardening is necessary at fainter flux levels. This can be confirmed
by future missions such as XMM and Chandra. On the other hand, at the
brighter flux range, $10^{-13} \sim 10^{-11}$ \ergs\ (2--10 keV) where
we need to cover large area to overcome the small surface number
densities, the AMSS provides the best opportunity for statistical
studies of X-ray sources in the universe.

\acknowledgments

We are grateful to Prof. H.~Inoue for stimulating discussion. We thank
members of the \asca\ team for their support in satellite operation and
data acquisition, and Dr. H.~Kubo for the help in analysis.

\begin{figure}[hpt]
\plotone{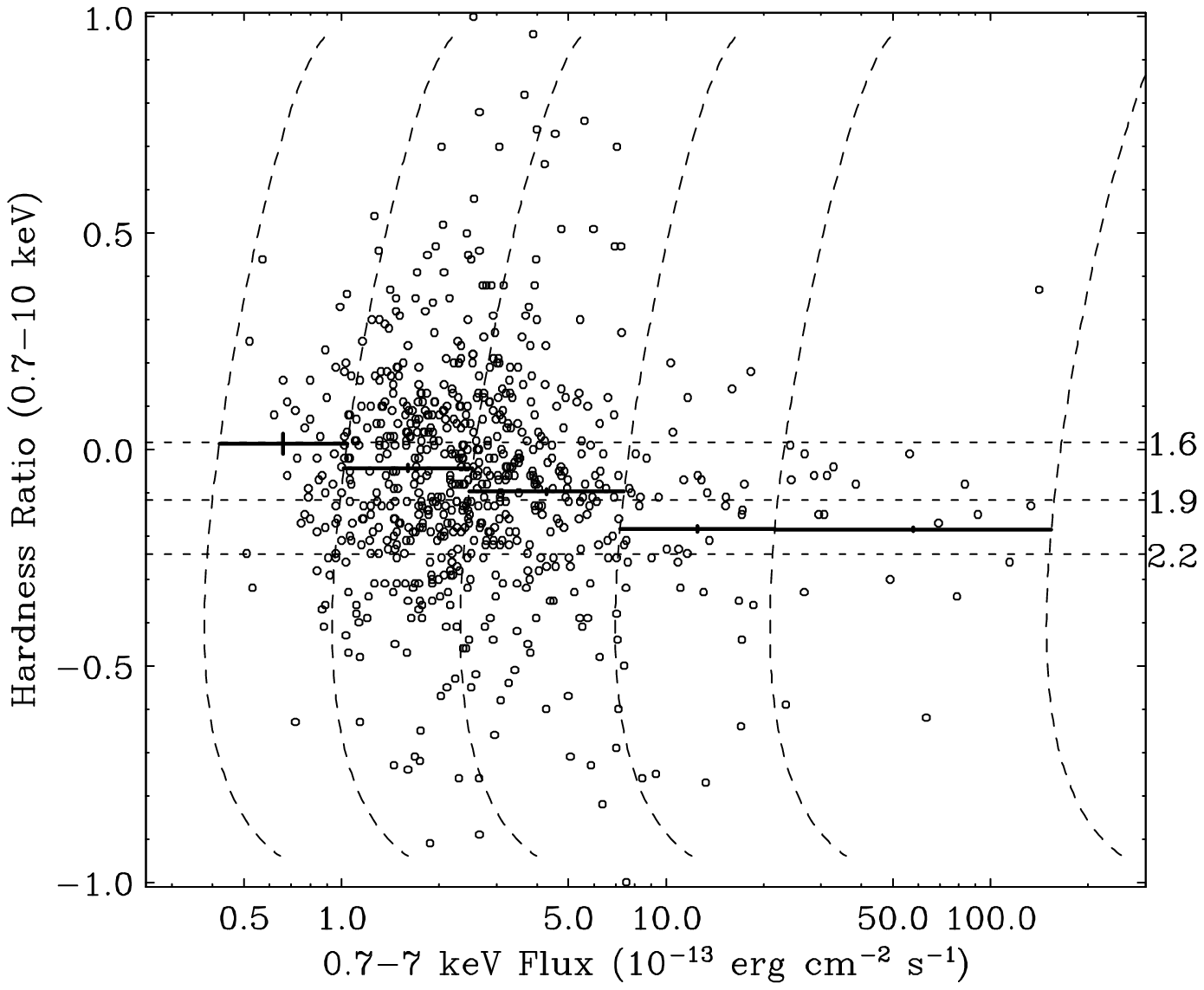}
\caption{
The correlation between the 0.7--7 keV flux and the hardness ratio
($HR1$) between the 0.7--2 keV and 2--10 keV count rate for sources
detected in the 0.7--7 keV survey in the AMSS sample. The count rate
is converted into flux for each source based on the best-fit hardness
assuming a power law spectrum and correcting for the Galactic
absorption. The crosses show the average hardness ratios (with
1$\sigma$ errors in the mean value) in the flux bin separated by the
the dashed curves, at which the count rate hence the sensitivity limit
is constant. The dotted lines represent the hardness ratios
corresponding to a photon index of 1.6, 1.9, and 2.2 assuming a power
law spectrum. Absorption with $N_{\rm H} = 3\times10^{21}$ cm$^{-2}$
increases $HR1$ by about 0.15.
\label{fig1}}
\end{figure}

\placefigure{fig1}

\begin{figure}[hpt]
\plotone{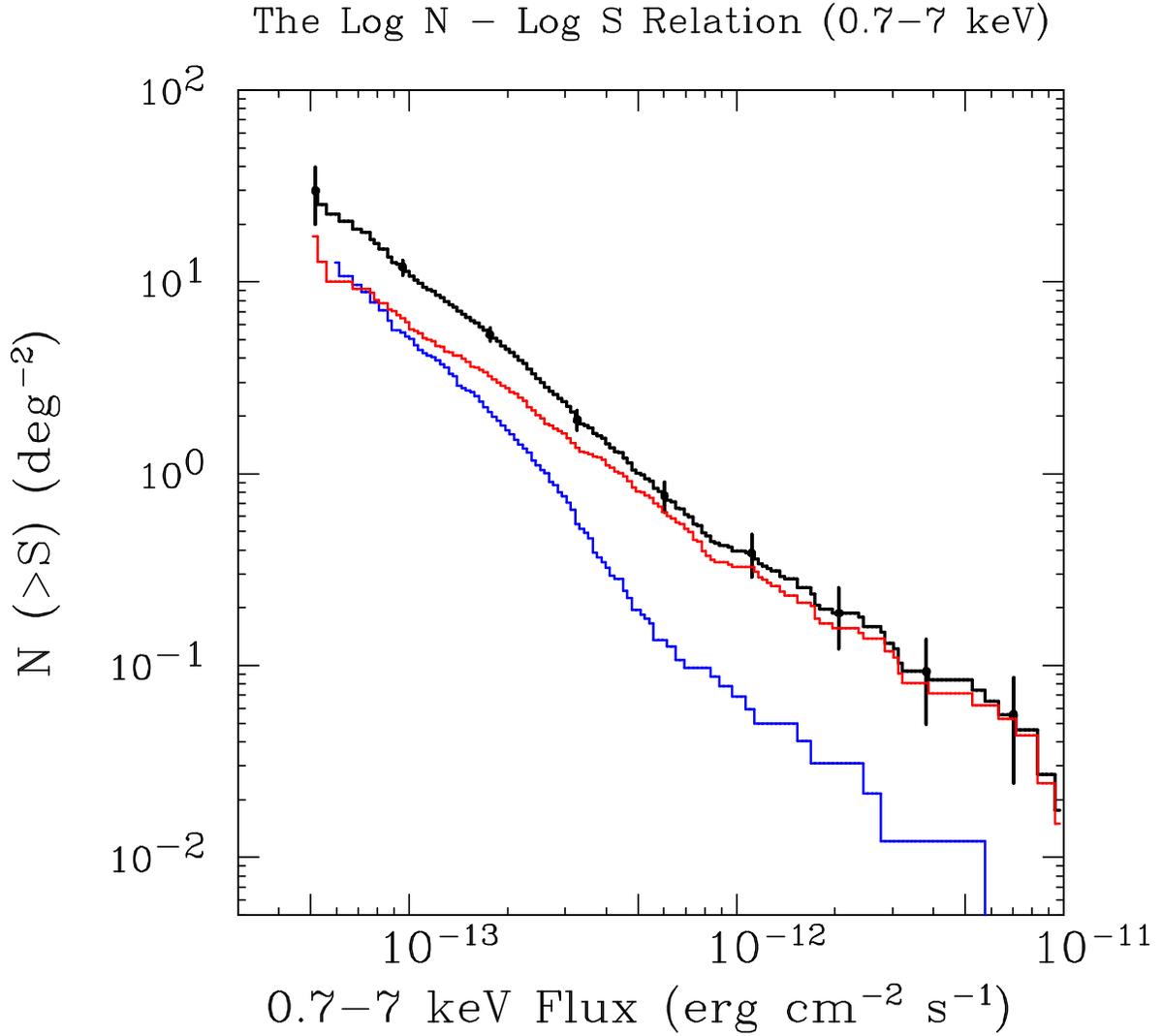}
\caption{
The integral \logn s in the 0.7--7 keV survey band, derived from the 
AMSS sample. The blue curve represents the result for the
hard source sample, consisting of sources with (apparent) photon
indices less than 1.7 in the 0.7--10 keV range, the red curve represents that 
for the soft source sample (with indices larger than
1.7), and the black curve represents the sum. At several points, the
90\% statistical errors in the source counts are plotted. A photon
index of 1.6 and 1.9 is assumed in the count rate to flux conversion
for the hard- and soft-source sample, respectively, with correction for
the Galactic absorption.
\label{fig2}}
\end{figure}

\placefigure{fig2}

\begin{figure}[hpt]
\plotone{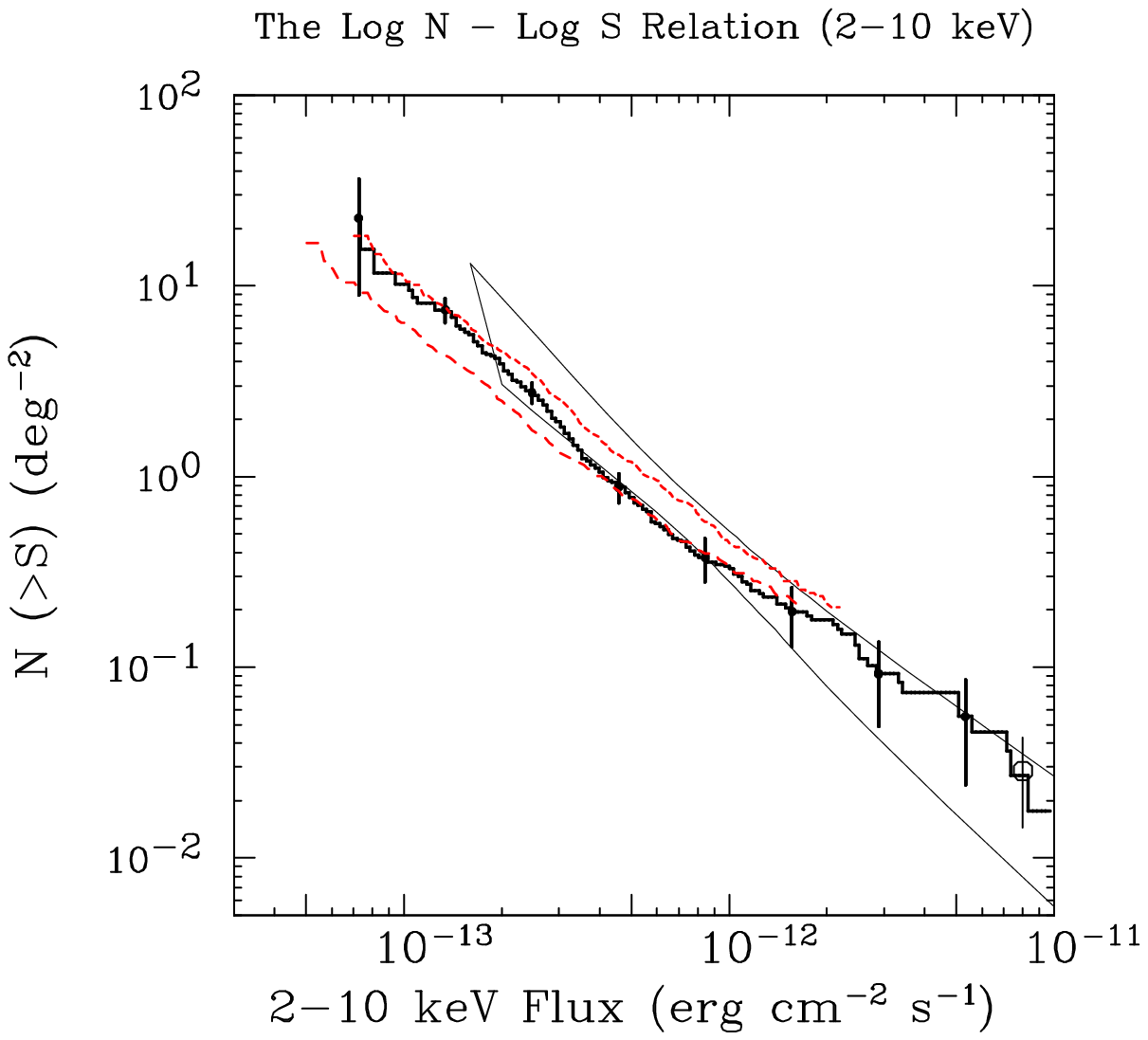}
\caption{
The integral \logn s in the 2--10 keV band survey, derived from the
AMSS sample. The elongated box represents the constraints by
the \ginga\ fluctuation analysis (Butcher \etal\ 1997). The open
circle corresponds to the source counts by \ginga\ survey (Kondo
\etal\ 1991). The two dashed curves (red) are the \logn s in the 0.7--2 keV 
band survey derived from the AMSS (this work), converted into the
2--10 keV flux assuming a power law with a photon index of 1.6
(short-dashed curve) and 1.9 (long-dashed curve).
\label{fig3}}
\end{figure}

\placefigure{fig3}

\end{document}